\documentclass{elsart}

\usepackage{psfig}

\usepackage{natbib}
\def\lax    {${_<\atop^{\sim}}$}
\def\gax    {${_>\atop^{\sim}}$}

\def\etal   {{\it et al.}~}

\begin{document}

\runauthor{S. Mathur}


\begin{frontmatter}

\title{New Insights  Into The Narrow-Line Seyfert 1 Phenomenon}

\author {Smita Mathur}
\address {Department of Astronomy, 140 West 18th Avenue, 
The Ohio State University, Columbus, Ohio 43210, USA}

\begin{abstract}

I briefly review the X/UV absorber models and show that the
observations of NLS1s are generally consistent with the models. The
covering factor of absorbers in NLS1s is likely to be high and there
is some evidence of super-solar metallicities. I argue that NLS1s may
be active galaxies in the early stage of their evolution and as such,
may be low luminosity, low redshift analogues of the high redshift
quasars. NLS1s may reside in rejuvenated gas-rich galaxies. I also
propose that the  high Fe~II emission in NLS1s may be a direct
consequence of their large accretion rate and so a collisional
ionization origin of Fe~II is favored.
\end{abstract}


\end{frontmatter}


\section{Introduction}

The organizers of this workshop have asked me to talk about the
X-ray/UV absorbers in AGN and discuss what they can tell us about the
Narrow Line Seyfert 1 phenomenon. However, I am going to go well beyond
my expertise and speculate. I think that this meeting is an ideal platform for 
discussing new ideas, so let me
start with my conclusions first and get them out of the
way. {\bf Conclusions:} 
\begin{enumerate}
\item The covering factor of warm absorbers is likely to be high in NLS1s.
\item Metallicities might be super solar.  
\item NLS1s may be  AGN in the
making and as such, may be low redshift, low luminosity analogues of
the high redshift quasars.  
\item NLS1s reside in rejuvenated/ gas rich
galaxies.
\end{enumerate}
Having said that, let's start from the beginning.

\subsection{The X/UV Absorbers}

The UV and X-ray intrinsic absorption systems observed in active
galactic nuclei (AGN) were never thought to be related to one another
because of their apparently different physical properties. The
inferred total column density of UV absorption lines was believed to
be about $10^{20}$ cm$^{-2}$, and the ionization parameter U was
thought to be such that CIV would be the dominant ionization phase of
carbon. In contrast, the cold X-ray absorbers are neutral with column
densities $> 10^{21}$ cm$^{-2}$. Models of warm X-ray absorbers also
required larger column densities and ionization parameters than the UV
absorbers. As was shown later, the ionization structure of the
absorbers was what was lacking from our understanding.

The above situation changed with the quasi-simultaneous 
observations of 3C351 with the ROSAT PSPC and the HST FOS. The UV spectrum of
3C351 showed OVI absorption line doublets and the X-ray spectrum
showed edges due to OVII/OVIII. These observations lead Mathur \etal
(1994) to conclude that the UV and X-ray absorbers are in fact one and
the same. The combined UV and X-ray analysis proved to be a powerful
technique to probe the physical properties of the nuclear regions of
AGN:  high column density, highly ionized, outflowing material
situated outside the broad emission line region. In addition, the
density of the absorber could be constrained in variable systems. This
was a previously unknown component  with important consequences:
an outflow that carries a significant amount of kinetic energy, and with a
mass outflow rate comparable to the accretion rate needed to power the
AGN engine.

Later work supported the unified X/UV absorber scenario (Mathur 1994,
Mathur \etal 1995, 1997, 1998, 1999, Shields \& Hamann 1997a, Crenshaw
1997). I will defer to Mathur (1997) for a discussion of controversies 
 regarding the X/UV models. I conclude that all the
observed data support this picture: the UV and X-ray absorbers are
physically related to each other. At the very least, the X-ray absorbers
make a substantial contribution to the absorption seen in the UV.

\section{X-ray \& UV Absorption in NLS1s}

Table 1 lists some NLS1 galaxies showing absorption in X-rays and/or
 UV. I have compiled this list from the literature and it is by no
 means complete. There are a few more ROSAT selected NLS1s with X-ray
 warm absorbers for which UV spectra are not available.

\begin{center}
\begin{table}[h]
\caption{NLS1s with X-ray and/or UV Absorbers }
\begin{tabular}{|lccc|}
\hline\hline
 Object & X-ray Warm  & UV  & Comments\\
                &  Absorber  &   Absorber          &      \\
\hline
IRAS 13349+2438 &   Y         &  Y (HST)    &  Only CIV, low S/N \\
Mrk 1298        &   Y         &  Y (IUE)    &  Low S/N in IUE   \\
I Zw 1          &   Y         &  Y (HST)    &  Excess in blue wing\\
    &             &             &  of UV emission lines.     \\
NGC4051         &   Y         &  N?         &   -do- \\
Mrk 507         &   Y         &  ?          &       \\
IRAS 17020+4544 &   Y         &  ?          &        \\
Q0117-2837      &  Y          &  N (IUE)    & Not expected  \\
    &             &             & (too ionized)                   \\
RX J 0134.3-4258  &   Y        &  ?         &  -do-  \\
PG 1404+266       &   Y/N       &  Y        &        \\
\hline
\end{tabular}
\end{table}
\end{center}

From Table 1 we see that for some NLS1s with X-ray warm absorbers
either (1) UV data do not exist, or (2) UV absorption is not expected
because the X-ray absorber is very highly ionized. I cannot comment on
the X/UV connection for these objects. Excluding these objects, we
have warm absorbers with either (1) UV absorption, or (2) blue
asymmetric emission lines. The first case is what is expected from 
the X/UV absorber models, while the second (blue asymmetric emission
lines) is a new observation and is intriguing. I would like to propose
that {\it the blue asymmetry is caused by emission from the
outflowing/inflowing warm absorber.} (See also B.\ Peterson's article
on the broad HeII emission line in NGC4051). A preliminary investigation
shows consistency with the predictions from photoionization models if
the covering factor is large. This leads to my first conclusion: {\bf
Warm absorbers in NLS1s have large covering factors.}

\begin{figure}[htb]
\centerline{
\psfig{figure=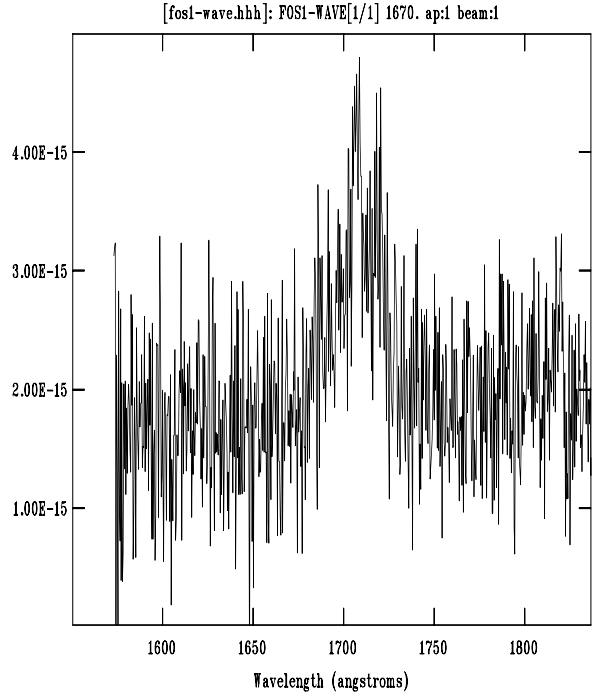,height=3.9truein,width=3.5truein,angle=0}
}
\caption{One orbit HST FOS spectrum of IRAS 13349+2438 showing associated CIV 
absorption. Wavelength in angstroms is plotted on the X-axis, while Y-axis 
corresponds to flux in units of ergs s$^{-1}$ cm$^{-2}$ \AA$^{-1}$.}
\end{figure}

Now consider the warm absorbers with UV
absorbers. The existence of UV absorbers is in itself favored by the
unified X/UV models. There are three such objects listed in Table 1:
(1) IRAS 13349+2438: the ROSAT spectrum of this object is discussed in
Brandt, Mathur, Reynolds \& Elvis (1997). Using photoionization models
and the parameters of warm absorbers, we can predict the column
density of UV absorption lines. HST FOS observed the CIV absorption line
in this object (Figure 1). Being a single orbit snapshot observation,
the signal-to-noise ratio is poor. Also, with just one line, the
result cannot be conclusive. Nevertheless, the observed CIV column
density is as predicted for a reasonable value of the b parameter, b$\sim
100$ km/s. (2) I Zw 1: This object has a weak warm absorber, and as
expected, a weak associated absorption line system of Ly$\alpha$, NV
and CIV was observed with HST (Laor  1998a). (3) Mrk 1298 (or
PG1126$-$041): CIV and NV absorption lines are clearly present in the
low dispersion, IUE spectrum of this object (Figure 2). The S/N,
however, is too low to develop detailed models. To conclude, the NLS1s
also support the unified X/UV absorber models.

This now brings me to the last object in the Table, PG1404+266.
\begin{figure}[htb]
\centerline{
\psfig{figure=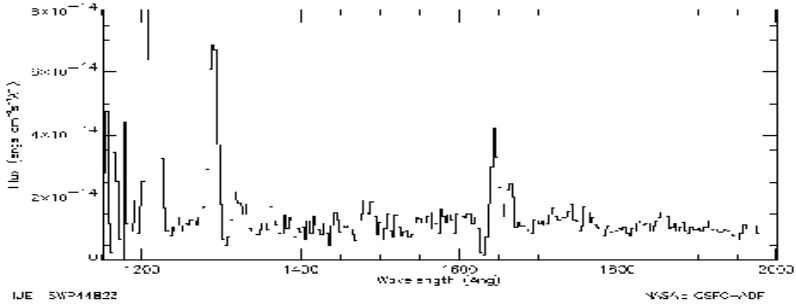,height=5.0truein,width=5.5truein,angle=0}
}
\vspace*{-2.5in}
\caption{IUE SWP Spectrum of PG1126-041 showing associated CIV and NV 
absorption lines (near 1640\AA\, and 1300\AA\, respectively).}
\end{figure}

\subsection{PG 1404+266}

ASCA, ROSAT and HST observations of PG1404+266 are discussed in Ulrich
\etal 1999. The X-ray spectrum of the object shows the signature of a
$\sim$1 keV absorption feature which may be interpreted as a
NeVII--NeX warm absorber (note that this would be different from the
OVII/OVIII warm absorbers generally observed with ROSAT). PG 1404+226
also has an unusually strong Fe K$\alpha$ emission line. The HST
spectrum shows the presence of the CIII$^{*}$\,$\lambda$1175.7 emission
line and associated absorption lines of Ly$\alpha$, NV and CIV. Ulrich
\etal found that, while the strength of the Ly$\alpha$ and CIV lines
was generally consistent with that expected from the warm absorber,
the strength of NV was not: the NV absorption line was {\it stronger} than
expected.

This is reminiscent of the metallicity determinations in quasars (see
review by Hamann \& Ferland 1999). As shown by Hamann \& Ferland,
nitrogen serves as an metallicity indicator. Moreover, nitrogen is
preferentially enhanced when metallicities are high, N$/$H $\propto$
Z$^2$. The observations of PG1404+226 thus suggest that {\bf nitrogen
is enhanced and so the metallicities are super solar} in this object
and likely to be in NLS1s in general. We are investigating whether this
suggestion is correct by performing a detailed metallicity analysis
(Mathur \& Komossa 2000).

\section{Are NLS1s Active Galaxies in the making?}

In addition to the case of PG 1404+226 discussed above, there are
  other lines of evidence suggestive of high metallicities in
  NLS1s. Wills \etal~(1999) found that the strength of the NV $\lambda
  1240$ emission line was systematically larger while the strength of
  CIV $\lambda 1549$ was systematically smaller in AGN with narrow
  emission lines (see also B.\ Wills, these proceedings). The strength
  of the fluorescent Fe-K alpha line in some NLS1s is also indicative
  of a super-solar abundance (A.\ Fabian, these proceedings). Thus the
  observations of emission as well as absorption lines in NLS1s imply
  super-solar gas phase metallicities.

 Such metal enrichment is possible when the initial mass function of
  star formation is flat, favorable for high mass star formation, and
  the evolution is fast. Such a star formation scenario is likely to
  be present in deep potential wells like galactic nuclei and
  protogalactic clumps (HF99). Moreover, high metallicities are
  achieved while consuming less gas (HF99). The NLS1s may then
  represent that early phase in galactic evolution when rapid star
  formation is taking place in the nucleus.

 Note also that NLS1s have relatively smaller BH masses. As per the well
known correlation of Magorrian \etal~(1998), smaller mass BHs reside in
galaxies with smaller spheroids. Since NLS1s have relatively smaller
mass BHs compared to normal Seyferts, the spheroids of their host
galaxies might be smaller (see also Laor 1998b). Indeed, in the
compilation of Wandel (1999), the NLS1 galaxy NGC4051 has the smallest
black hole to bulge mass ratio. An accreting BH would also grow in
mass with time [the Salpeter time scale of growth is determined by
t$_s = 3 \times 10^7 (L_{Edd}/L_B) \eta_{0.1}$ yr.\ where $\eta_{0.1}$
is the radiative efficiency in units of 0.1 (see Fabian
1999)]. Since NLS1s accrete at close to the Eddington limit, their BHs
would grow faster. So, smaller BHs in NLS1s are likely to be younger
as well. 

 These arguments support my proposal that {\bf 
NLS1s might be Active Galaxies in early phases of their evolution.}

\section{Similarity between NLS1s and the high redshift quasars.}

Hamann \& Ferland (1993) found high metallicities in high redshift
 quasars (Z\gax Z$_{\odot}$ at z\gax 4). Similarly we find that NLS1s
 may also have large metallicities.

Similarities between the observed properties of low ionization Broad
  Absorption Line Quasars (BALQSOs) and NLS1s have been reported in
  the literature (e.g. Lawrence \etal~1997, Leighly \etal~1997). Both
  these classes show strong Fe~II$\lambda 4570$ and AlIII$\lambda 1857$
  and weak CIV$\lambda 1549$ and [OIII]$5007$ emission lines. Their
  continua are red in the optical and strong in the IR. Evidence of
  relativistic outflow is also reported in three NLS1s (Leighly
  \etal~1997). If these two classes are indeed related (see also
  Brandt, these proceedings), then NLS1s, at least those with
  some evidence of outflow, might be low redshift, low luminosity
  cousins of BALQSOs. BALQSOs are tentatively identified with a 
  phase in quasar evolution when the matter around the nuclear BH is
  being blown away, and a quasar emerges (see, e.g.\ Fabian
  1999). NLS1s may then represent a similar early evolutionary phase
  at low redshift.

Optical spectra of a sample of z\gax 4 quasars revealed that their
  emission lines are typically narrower than the low redshift quasars
  (FWHM \lax 2000 km/s, Shields \& Hamann 1997b). The normal
  explanation of this observation is that these are type 2 quasars,
  where the broad emission lines are obscured from our line of
  sight. Alternatively, these high z quasars might be true ``narrow''
  broad line objects (see Mathur 2000 for a discussion of selection
  effects).
 
There is  another interesting connection with high redshift. As
 discussed above, NLS1s have strong Fe~II emission lines.  Quasars
 Q0014+813 and Q0636+680 at redshifts z=3.398 and z=3.195 respectively,
 were observed to have very strong Fe~II emission (Elston, Thompson, \&
 Hill 1994). Are they also highly accreting objects at an early
 evolutionary phase? Note also the narrow UV emission lines (FWHM \lax
 2150 km s$^{-1}$) in the ultra strong UV Fe~II emitter Q2226-3905
 (Graham, Clowes, \& Campusano, 1996).

All these similarities point towards {\bf NLS1s being low redshift, low
 luminosity analogues of high redshift quasars.}

\section{Do NLS1s reside in rejuvenated galaxies?}

I have argued that NLS1s may represent an early phase in AGN
 evolution. Whether they reside in young galaxies is a separate
 question and a step further. That young galaxies are gas rich is
 helpful; they would have the large reservoir of gas necessary to
 sustain the close to Eddington-rate accretion in NLS1s. But do we
 have any evidence that they indeed reside in young galaxies? There is
 no published systematic study of the properties of the host galaxies
 of NLS1s. However, some of the NLS1s are originally from the Zwicky
 (e.g.\ I Zw 1) and Markarian (e.g.\ Mrk 766) samples of galaxies
 implying that they are blue. While the blue color might be due to big
 blue bumps in the active nuclei, as in normal Seyfert galaxies, NLS1s
 have weak blue bumps and so the blue colors might be a result of
 actively star forming galaxies. Some NLS1s (e.g.\ IRAS 13349+2438),
 are infrared luminous, and star forming.  Using the galaxy catalogs
 RC3 (de Vaucouleurs \etal~1991) and UGC (Nilson 1973), we
 looked into the morphology of a small sample of NLS1s listed in Table
 1 and found information on seven of them. Three were found to be
 compact (I Zw 1, Mrk 507, and Mrk 1298), two showed signatures of an inner
 ring (NGC 4051 and Ark 564), and three have nuclear bars (NGC 4051,
 Mrk 766 and Ark 564).  These are signatures of recent activity
 (though not necessarily), quite likely due to galaxy-galaxy
 interactions or mergers.  In this scenario the galaxies are newly
 formed, or rejuvenated.

That NLS1s reside in young galaxies is also consistent with the
 hypothesis that the formation and evolution of galaxies and their
 active nuclei is intimately related (Rees 1997, Fabian 1999, Granato
 \etal 1999, Haehnelt \& Kauffmann 1999). In this scenario, the process
 of formation of a massive BH and the active nucleus is the very
 process of galaxy formation. The active nucleus and the galaxy evolve
 together, with the BH accreting matter and the galaxy making stars. At
 one stage the winds from the active nucleus blow away the matter
 surrounding it and a quasar emerges. This is not only the end of the
 active evolution of the quasar, but of the galaxy as well, as it 
 is evacuated of its interstellar medium. The quasar then shines as
 long as there is fuel in the accretion disk (Fabian 1999). In this
 scenario, high redshift quasars represent an early stage of galaxy
 evolution, BALQSOs at z$\approx 2$ represent the stage when the gas
 is being blown away, and z$\approx 1$ quasars would be the passively
 evolving population. Massive ellipticals found today might be the
 dead remnants of what were once quasars.

The quasar phenomenon may thus be a result of galaxy formation due to
 primordial density fluctuations. At low redshift, when new galaxies
 are formed due to interactions or mergers, similar evolution may take
 place. As argued above, the NLS1s may represent a crucial early
 phase. (In our scenario, the accretion rate \.m = \.M/\.M$_{Edd}$ is
 large in the early stages of evolution and reduces later on. This is
 opposite to the proposal by Wandel (1999) in which \.m increases with
 time.)

 In fact, there might be some NLS1s with a starburst component. Soft
 X-ray spectra of NLS1 are steep and often variable. However, Leighly
 \etal~(1996) and Page \etal~(1999) report that while the power-law
 component in the NLS1 Mrk 766 varied, the thermal black-body
 component did not. This component might well be due to a nuclear
 starburst. Note also the strong CO emission in the prototype NLS1 I
 Zw 1 (Barvainis, Alloin \& Antonucci 1989). Schinnerer, Eckart \&
 Tacconi (1998) mapped I Zw 1 in CO and found a circumnuclear ring of
 diameter 1.8 kpc. The authors found strong evidence for a nuclear
 starburst. There is also a companion to I Zw 1, supporting an
 interpretation of the starburst being due to an interaction. Similarly, AGN
 activity is known to exist in starburst galaxies (see Heckman 1999
 for a review). The poster by Dennefeld \etal~ reports observations of
 narrow optical emission lines in a sample of IR selected starburst
 galaxies. See Mathur (2000) for the details of the connection between
 NLS1s and the evolution of galaxies and active galaxies.

\section{On strong Fe~II emission in NLS1s.}

There is a general consensus that large accretion rate, \.m, is a
 possible driver of many of the observed properties of NLS1s.
 Strength of Fe~II emission may also be linked to the large accretion
 rate. In a model by Kwan \etal~(1995), Fe~II line emission is produced
 in an accretion disk.  The accretion disks with larger accretion rate
 may simply have more mass to produce stronger Fe~II. Thus I would like
 to argue in favor of collisional ionization as the origin of the
 Fe~II emission in NLS1s and AGNs in general (see also the article by
 S. Collin in these proceedings). Remember also the
 CIII$^{*}$\,$\lambda$1174.7 observed in PG 1404+266. This line is
 also seen in the HST spectrum of another NLS1, I\,Zw\,1 (Laor \etal
 1997). Laor \etal preferred the explanation that it is produced by
 resonance scattering of continuum photons. However, in this scenario
 the velocity gradient in each emitting cloud and the total covering
 factor would be exceptionally large. The presence of the 
 CIII$^{*}$\,$\lambda$1174.7 line, together with strong Fe~II emission,
 suggests a collisional origin.

\section{Conclusions}

 That the AGN phenomenon was so much stronger at z$\sim$2--3 than
 today has long elicited the suspicion that there is a connection
 between the youth of a galaxy and the likelihood that an AGN forms
 inside it. The question then naturally arises, ``what are the local
 counterparts to the young galaxies in the early universe in which
 local AGN may live?'' (Krolik 1999). A standard answer to
 this question is ``Starburst galaxies''. Heckman (1999) has argued
 that starburst galaxies are the low redshift analogues of Lyman break
 galaxies at high redshift. Similarly, we ask, what are the low
 redshift analogues of high redshift (z\gax 4) quasars? I propose
 that they might be NLS1s.

 It is my pleasure to thank Th. Boller and the organizing committee
 for inviting me to this wonderfully stimulating workshop. I thank the
 Wilhelm and Else Heraeus foundation for travel support and the
 delightful stay at the Physikzentrum, Bad Honnef. This work is
 supported in part through NASA grant NAG 5-3249 (LTSA).
The figures were  created using the archives at the Space Telescope Science
 Institute, operated by the Association of Universities for
 Research
 in Astronomy, Inc., from NASA contract NAS5-26555.





\end{document}